\documentclass[prd,nofootinbib,onecolumn,english,cp1251,T2A,showpacs,showkeys]{revtex4}
\usepackage{babel}

\usepackage{amssymb}
\usepackage{amsmath}
\usepackage{slashed}

\usepackage[dvips]{graphicx} 

\usepackage{amsthm} 

\usepackage{varioref}   
\usepackage{xr}         

\usepackage{pstricks}           

\allowdisplaybreaks[1]

\newcommand\ba{\begin{eqnarray}}
\newcommand\ea{\end{eqnarray}}
\newcommand\baf{\begin{eqnarray}}   
\newcommand\eaf{\end{eqnarray}}

\begin{document}

\title{High Energy Hadron Spin Flip Amplitude
 \footnote{Talk at the workshop
  "Selected  problems in
quantum field theory", Dubna (2015)
dedicated to the memory of
professor E.A. Kuraev} }

\author{O.V. Selyugin}
\email{selugin@theor.jinr.ru}
\affiliation{Joint Institute for Nuclear Research, 141980 Dubna, Moscow Region, Russia}

\date{\today}

\begin{abstract}
 The high energy part of the hadron spin flip amplitude is examined in
    the framework of the new high energy general structure (HEGS) model
    of the elastic hadron scattering at high energies.
    The different forms of the hadron spin flip amplitude are compared in the
    impact parameters representation.
   It is shown that the existing experimental data of the proton-proton and proton-antiproton
 elastic scattering at high energy in the region of the diffraction minimum and at
  large momentum transfer give support in the presence of the energy-independent part
  of the hadron spin flip amplitude with the momentum dependence proposed in the
  works by Galynskii-Kuraev.
\end{abstract}



\maketitle

\section{Introduction}

 The spin effects play very often the touchstone role for  many different theoretical approaches. Especially, it applies to
     the  hadron-hadron elastic processes.
     The famous experiments carried out by A. Krish at the ZGS  to obtain the
     spin-dependent differential cross sections \cite{Krish1a,Krish1b} and
     the spin correlation parameter $A_{NN}$   \cite{Krish1c}
     and   at the AGS \cite{Krish2} to obtain  the spin correlation parameter $A_{N}$
      showed the significant spin
     effects at large momentum transfer.
         Note that the spin-flip amplitudes, determined by the non-leading Reggions exchange,
         give the large contribution only at small momentum transfer, as the massive non-leading Reggions have a large slope         and  decrease faster at large $t$.
     Of course,  many questions arise about the
     energy dependence of the spin flip amplitudes and the relative phase between the spin non-flip and
     spin-flip amplitudes.
     Now there are many different models for the description of the elastic
  hadrons scattering amplitude \cite{mog1}.
  They lead to  different predictions of the structure of the scattering
  amplitude at high energies.
  The diffraction  processes at  very high
 energies, especially in the  TeV energy region, are simplified by the asymptotical
  regime but can display complicated features    \cite{FS-07,dif04}.

According to the standard opinion, the hadron spin-flip amplitude is
  connected with the quark exchange between the scattering hadrons,
  and at large energy it can be neglected.
  Some models, which take into account the non-perturbative effects,
  lead to the non-dying hadron spin-flip amplitude \cite{mog2,GKS,Dor-Koch,Anselmino}.
  Another complicated question is related to the difference
  in phases of the spin-non-flip and spin-flip amplitude.

       The description of the high energy processes requires  using some
     unitarization procedures of the Born  scattering amplitude.
      They are related to the different forms of the summation of some sets of diagrams of the tree approximation.
    The unitarization leads to the asymptotic unitarity bound
   connected with the so-called  Black Disk Limit (BDL), which can leads to different saturation effects \cite{BDL}.
    In the partial wave language, we need to sum many different waves with
    $l \rightarrow \infty $ and this leads to the impact parameter representation \cite{Predazzi66} converting
    the summation over $l$  into  integration over $b$.

\section{Spin dependent  scattering amplitude }

      In the  Regge limit $t_{fix.}$ and $s \rightarrow \infty$  one can write the
       Regge-pole contributions to  the helicity amplitudes in the $s$-channel as
       \cite{Capella}
\begin{eqnarray}
    \Phi^{B}_{\lambda_1, \lambda_2,\lambda_3, \lambda_{4}} (s,t) \sim
    \sum_{i} g^{i}_{\lambda_1, \lambda_2}(t)  g^{i}_{\lambda_3, \lambda_4}(t)
    [\sqrt{|t|}]^{|\lambda_1- \lambda_2|+|\lambda_3- \lambda_4|}
    (\frac{s}{s_0})^{\alpha_{i}} (1 \pm e^{-i \pi \alpha_{i} }).
\end{eqnarray}
  A convenient basis is in terms of helicities, where the corresponding
  amplitudes of proton-proton scattering are in the $s$-chanel \cite{Lehar}, is
  \begin{eqnarray}
    \Phi^{B}_{1} (s,t) &=& <++|++>; \ \ \  \Phi^{B}_{2} (s,t) = <++|-->; \nonumber \\
   \Phi^{B}_{3} (s,t) &=& <+-|+->; \ \ \  \Phi^{B}_{4} (s,t) = <+-|-+>; \ \ \
   \Phi^{B}_{5} (s,t) = <++|+->;
\end{eqnarray}

 The differential cross section is
 \begin{eqnarray}
   \frac{d \sigma}{dt} = \frac{2 \pi}{s^2} ( |\Phi_{1}|^2 + |\Phi_{2}|^2+
   |\Phi_{3}|^2 + |\Phi_{4}|^2+4 |\Phi_{5}|^2.
\end{eqnarray}
  The total helicity amplitudes can be written as
   \begin{eqnarray}
  \Phi_{i}(s,t) =
  \Phi^{h}_{i}(s,t)+\Phi^{em}_{i}(s,t) e^{\varphi(s,t)},
  \end{eqnarray}
   where
$\Phi^{h}_{i}(s,t)$ is the pure strong interaction of hadrons,
$\Phi^{em}_{i}(s,t)$ is the electromagnetic interaction of hadrons and
$\varphi{s,t}$
is the electromagnetic-hadron interference phase factor \cite{WY,Selphase}.
The  corresponding  spin-correlation values will be
\begin{eqnarray}
  A_{N} \frac{d \sigma}{dt} = -\frac{4 \pi}{s^2}
   [ Im (\Phi_{1}(s,t) + \Phi_{2}(s,t)+ \Phi_{3}(s,t) - \Phi_{4})(s,t) \Phi^{*}_{5}(s,t)]
\end{eqnarray}
and
\begin{eqnarray}
  A_{NN} \frac{d \sigma}{dt} = \frac{4 \pi}{s^2}
   [ Re (\Phi_{1}(s,t) \Phi^{*}_{2}(s,t) - \Phi_{3}(s,t) \Phi^{*}_{4})(s,t) + |\Phi_{5}(s,t)|^2]
\end{eqnarray}

 The $s$-channel factorization together with the experimental information
 about the spin-correlations effects at  high energy and small momentum transfer
 in the proton-proton  elastic scattering suggests  that the double
 helicity flip as a second-order effect and, consequently, the amplitudes $\Phi^B_{++--} (s,t)$ and  $\Phi^B_{+--+} (s,t)$   can be  neglected.
 Furthermore, when the exchange of Regge poles has natural parity, we have
 for the spin-non-flip amplitudes \cite{Morel}
  $\Phi^B_{++++} (s,t) = \Phi^B_{+-+-} (s,t)$

   Neglecting  the $ \Phi_{2}(s,t)- \Phi_{4}(s,t)$ contribution the spin correlation parameter $A_{N}(s,t)$
   can be written taking into account the phases of  separate amplitudes
  \begin{eqnarray}
  A_{N}(s,t)  \frac{d \sigma}{dt} = -\frac{4 \pi}{s^2}
   [ (F_{nf}(s,t)| \ |F_{sf}(s,t)| Sin(\phi_{nf}(s,t)-\phi_{sf}(s,t)).
\end{eqnarray}
    where  $\phi_{nf}(s,t), \phi_{sf}(s,t)$ are the phases of the spin non-flip and spin-flip amplitudes.
  It is clearly seen that despite the large spin-flip amplitude the analyzing power can be near zero
  if the difference of the phases is zero in some region of momentum transfer.
  The experimental data at some point of the momentum transfer show the energy independence of
  the size of the spin correlation parameter $A_{N}(s,t)$ (see Fig.1).
  Hence, the small value of the $A_{N}(s,t)$ at some $t$ (for example, very small $t$)
  does not serve as a proof that it will be small in other regions of momentum transfer.

\begin{figure}
\includegraphics[width=.45\textwidth]{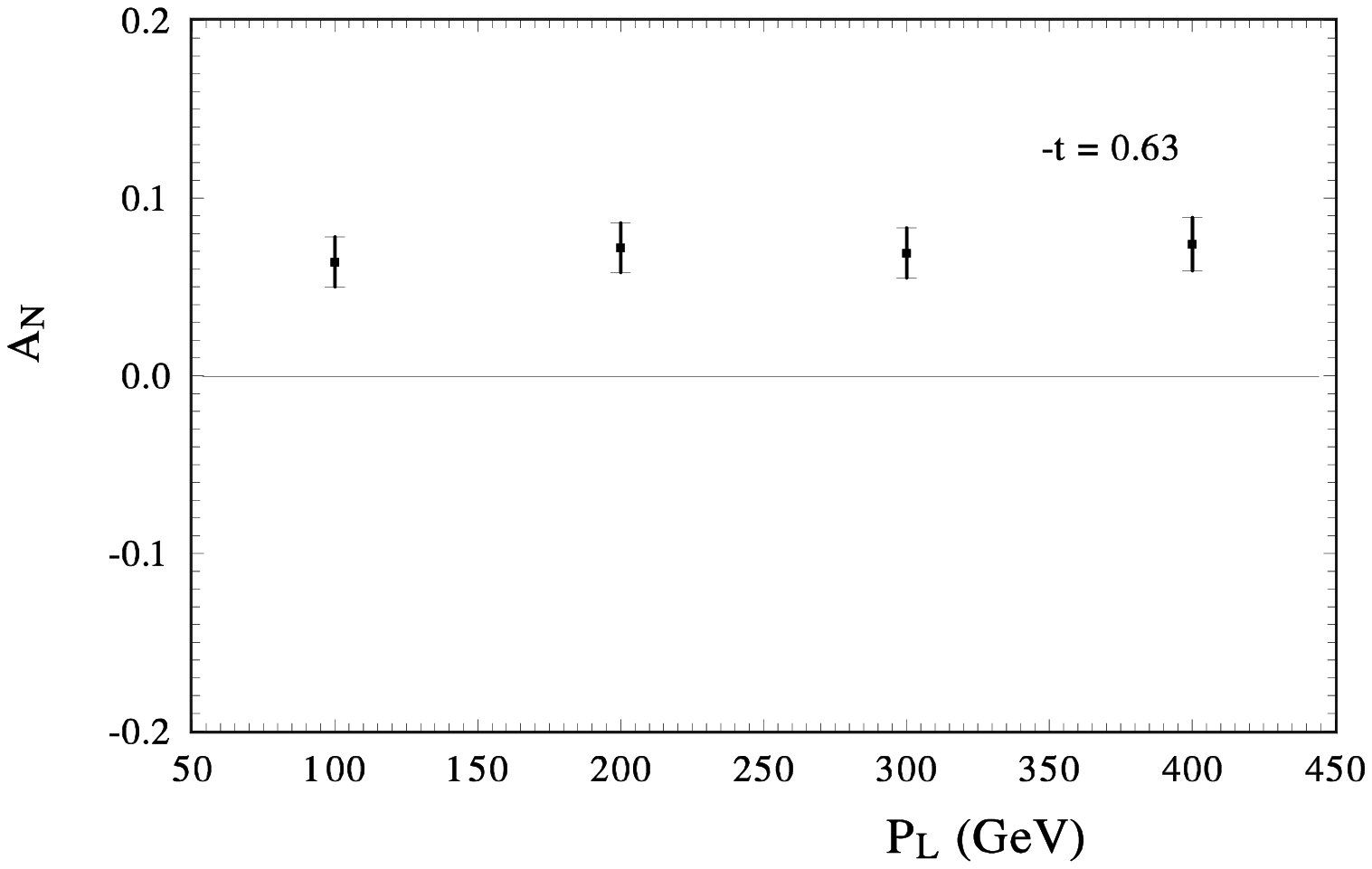} 
\includegraphics[width=.45\textwidth]{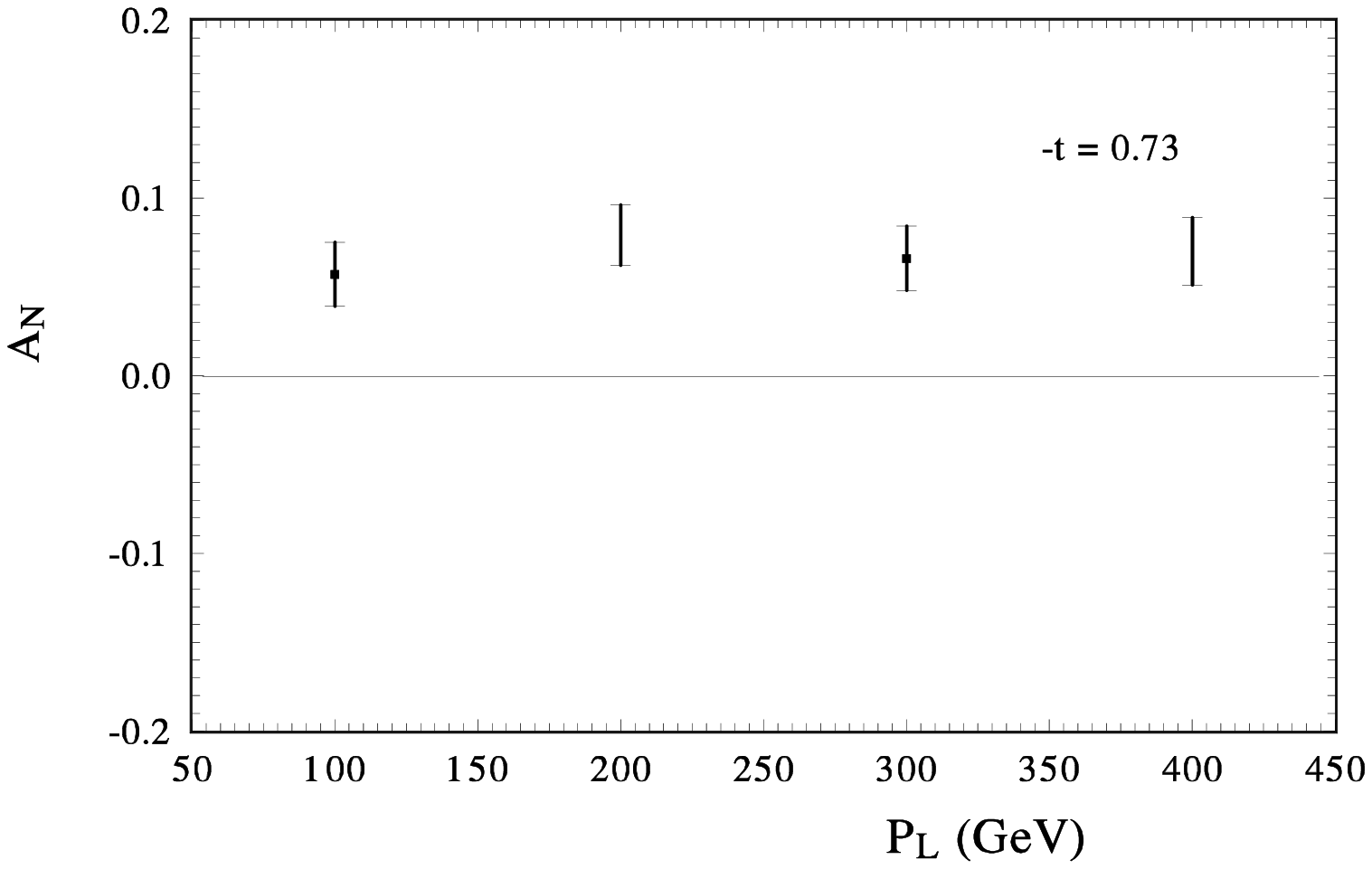} 
\caption{  The energy dependence of the experimental data of the spin correlation parameter $A_{N}(s,t)$
 at some separate point of the momentum transfer \cite{Spires-data} \\
 a) [left] at $t=0.63$ GeV$^2$ and b) [right]  at $t=0.63$ GeV$^2$
  }
\label{Fig_1}
\end{figure}

\section{Elastic nucleon scattering (HEGS model) }

      In some early models  of the elastic hadron scattering the hadron was regard as
      the whole particle  and the diffraction occurred on the surface \cite{Wu,Shremp} of the hadron.
  The parton structure of the hadron can be related to the hadron form factors and the elastic scattering amplitude
  is proportional to these form factors \cite{VanHove}.
  Some models take into account the parton structure of the hadron as
     separate interactions of the gluons and quarks \cite{Block}.
  A more complicated picture of the hadron structure appears with introducing
  the non-forward parton structure - general parton distributions (GPDs) and transfer momentum
   distributions (TMD). The different moments of  GPDs allow us  to calculate the different
   hadron form factors, such as Compton form factors ($R_{V}(t)$, $R_{T}(t)$, $R_{A}(t)$),
   electromagnetic form factors ($F_{1}(t)$, $F_{2}(t)$,
   and so-called  gravimagnetic form factors
   (  $A(t)$, $B(t)$  \cite{Ji97,Kroll04,D15}. 

  Let us use the obtained momentum transfer dependence of GPDs to calculate the electromagnetic and gravimagnetic form factors
  of the nucleons as the first and second moments of GPDs  \cite{ST-PRDGPD,GPD-PRD14}.
  The obtained form factors are related to the charge and matter distributions.
  By fixing the parameters of the obtained form factors
  the new High Energy General Structure (HEGS) model of the elastic proton-proton and proton-antiproton scattering
  was proposed \cite{HEGS0,HEGS1}.

 The  (HEGS)  model \cite{HEGS1}  gives a quantitative     description of the elastic nucleon scattering
 at high energy  with only $5$ fitting high energy parameters.
  A successful description  of the existing experimental data by the model shows that
   the elastic scattering  is determined by the generalized structure of the hadron.
The model leads to a  coincidence of the model calculations with the preliminary data at 8 TeV.
   We found that the  standard eikonal approximation \cite{Unit-PRD}   works perfectly well from
  $\sqrt{s}=9$  GeV up to  $\sqrt{s}=8$ TeV.

  In the model the Born term of the elastic hadron amplitude is determined
  \begin{eqnarray}
 F_{h}^{Born}(s,t) \ =  h_1 \ F_{1}^{2}(t) \ F_{a}(s,t) \ (1+r_1/\hat{s}^{0.5}) 
    \    +  h_{2} \  A^{2}(t) \ F_{b}(s,t) \ (1+r_2/\hat{s}^{0.5}),
    \label{FB}
\end{eqnarray}
  where $F_{a}(s,t)$ and $F_{b}(s,t)$  has the standard Regge form 
  \begin{eqnarray}
 F_{a}(s,t) \ = \hat{s}^{\epsilon_1} \ e^{B(s) \ t}; \ \ \
 F_{b}(s,t) \ = \hat{s}^{\epsilon_1} \ e^{B(s)/4 \ t}.
\label{FB-ab}
\end{eqnarray}
  The form factors $F_{1}(t)$ and $A(t)$ are determined by the first and second moments of GPDs,
  respectively,  and reflect the charge and matter distributions.

   The model takes into account the Odderon contribution with factor
   $h_{odd} = i h_{3} t/(1-r_{0}^{2} t) $. So the full Born term of the
   scattering amplitude is
 \begin{eqnarray}
 F_{h}^{Born}(s,t)=&&h_1 \ F_{1}^{2}(t) \ F_{a}(s,t) \ (1+r_1/\sqrt{\hat{s}})  
     +  h_{2} \  A^{2}(t) \ F_{b}(s,t) \    \\ \nonumber
    && \pm h_{odd} \  A^{2}(t)F_{b}(s,t)\ (1+r_2/\sqrt{\hat{s}}),
    \label{FB}
\end{eqnarray}
  The terms  proportional to $r_1/\sqrt{\hat{s}}$ and $r_2/\sqrt{\hat{s}}$ take into account
  some possible non-asymptotic contributions.

     The size and the energy and momentum transfer dependence of the real part of the elastic scattering amplitude are determined by the complex energy
     $\hat{s}=s \ exp(-i\pi/2)$. Hence, the model does not introduce some special functions or assumptions
     for the real part of the scattering amplitude.
     Note that the role of the real part
     is especially important at low momentum transfer (where the differential cross sections are
     determined by the Coulomb-hadron interference effects) and in the region of the diffraction minimum
     (where the imaginary part of the scattering amplitude has the zero,
      and the size of the diffraction minimum
     is determined by the real part of the scattering amplitude and the contribution of the
     spin-flip part of the scattering amplitude).

      The hard pomeron contribution was analyzed in the framework of the  model    \cite{NP-HP}.
    It was shown that such a contribution is very small in the elastic hadron scattering
    and is not felt in the fitting procedure.

  The final elastic  hadron scattering amplitude is obtained after unitarization of the  Born term.
    So, first, we have to calculate the eikonal phase.
 in the most favorable  eikonal unitarization scheme
   the eikonal phase corresponds to the Born term of the  scattering amplitude
   eq.(1) and in the  common case corresponds to
   the spin-dependent potential.

\begin{figure}
\includegraphics[width=.8\textwidth]{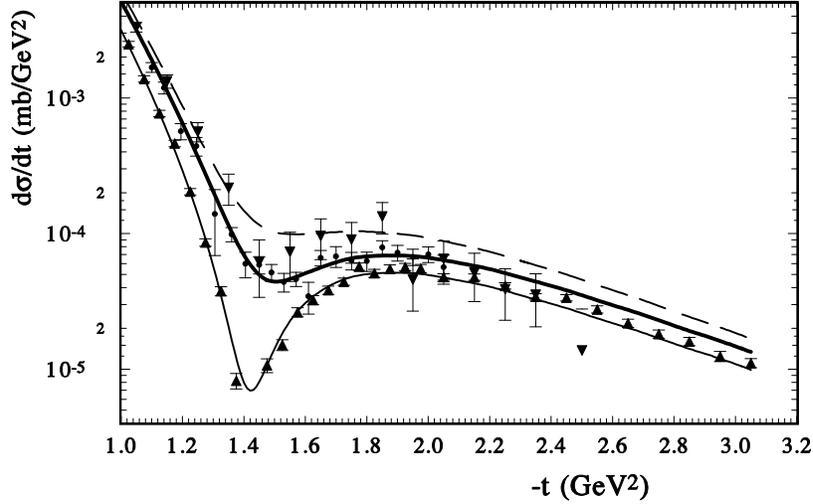} 
\caption{ The energy dependence of the diffraction minimum.
  The model calculations: dashed line  - $\sqrt{s} =9.78$ GeV; thick hard line - $\sqrt{s} =19.4$ GeV;
  thin hard line - $\sqrt{s} =30.5$ GeV; Triangles down, circles and triangles up - the experimental data
  at the same energies, respectively.
  }
\label{Fig_14}
\end{figure}

     The model description of the differential cross sections in such  a huge energy region
 is reflected  brightly  in the description of the region of the diffraction minimum (see Fig.2).
   In most part, the form and energy dependence of the diffraction  region is determined by the real part of the
   scattering amplitude and possible contributions of the spin-flip amplitude.
   Note that in the HEGS model only the Born term is determined
   and the complicated diffractive
   structure is obtained only after the unitarization procedure.

\section{Impact of the spin-flip contribution}

  Usually, one makes the
  assumptions that the imaginary and real parts of the spin-non-flip
  amplitude have the exponential behavior with the same slope, and the
  imaginary and real parts of the spin-flip amplitudes, without the
  kinematic factor $\sqrt{|t|}$ \cite{sum-L},
   are proportional to the corresponding parts of the non-flip amplitude.
 For example, in \cite{akth}
  the spin-flip amplitude was   chosen in  the form
 \ba
       F_{h}^{fl}=\sqrt{-t}/m_{p} \ h_{sf} \ F_{h}^{nf}.
       \label{sflip}
\ea
  That is not so as regards the $t$ dependence
  shown in Ref. \cite{soff}, where
  $F^{fl}_{h}$ is multiplied 
  by the special function
   dependent on $t$.
  Moreover, one mostly  takes  the energy independence of
  the ratio of the spin-flip parts to the spin-non-flip parts of the
  scattering amplitude.                         
All this is our theoretical uncertainty \cite{Cudpred-EPJA,M-Pred}.

\begin{figure}
\includegraphics[width=.45\textwidth]{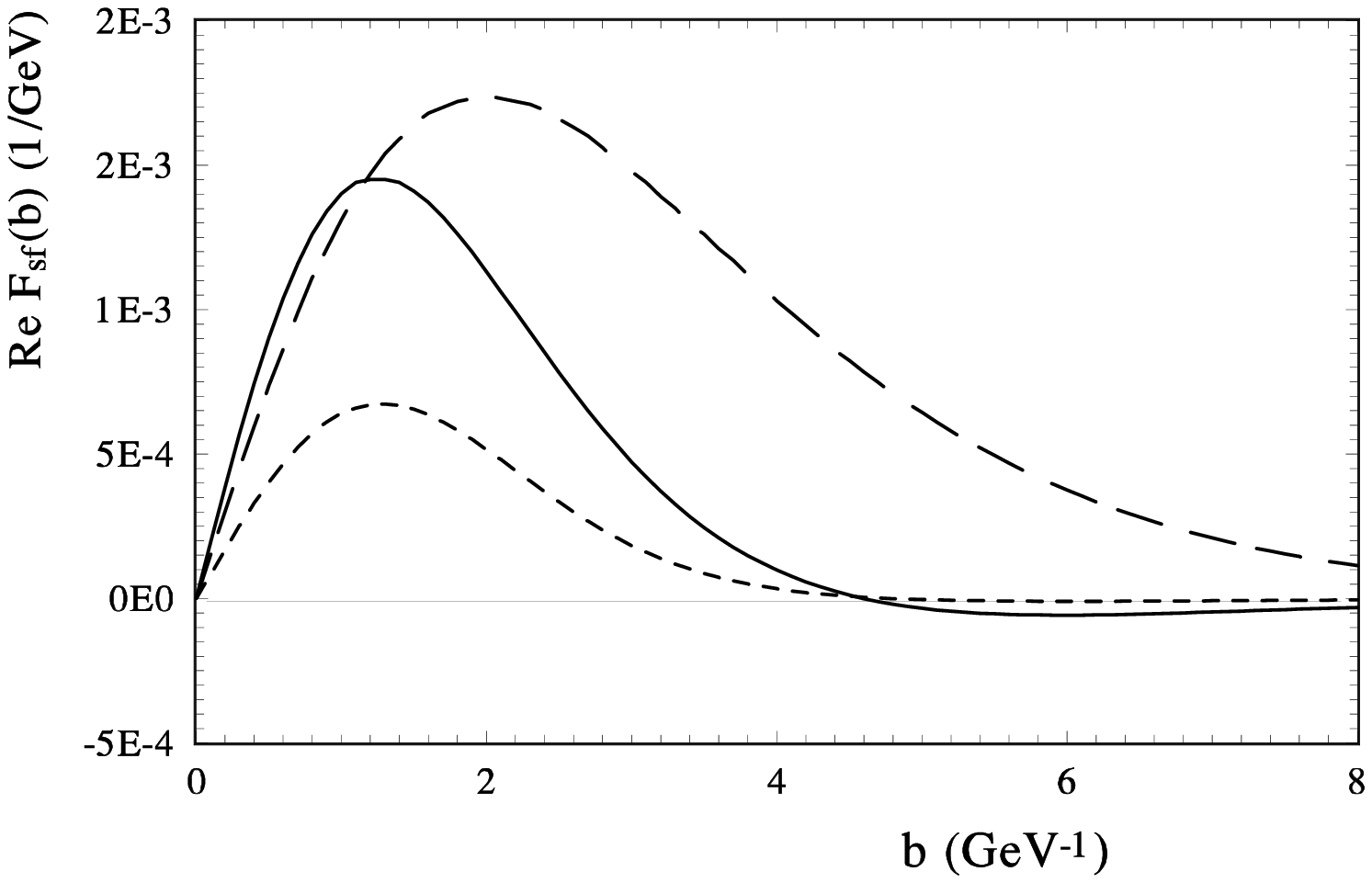} 
\includegraphics[width=.45\textwidth]{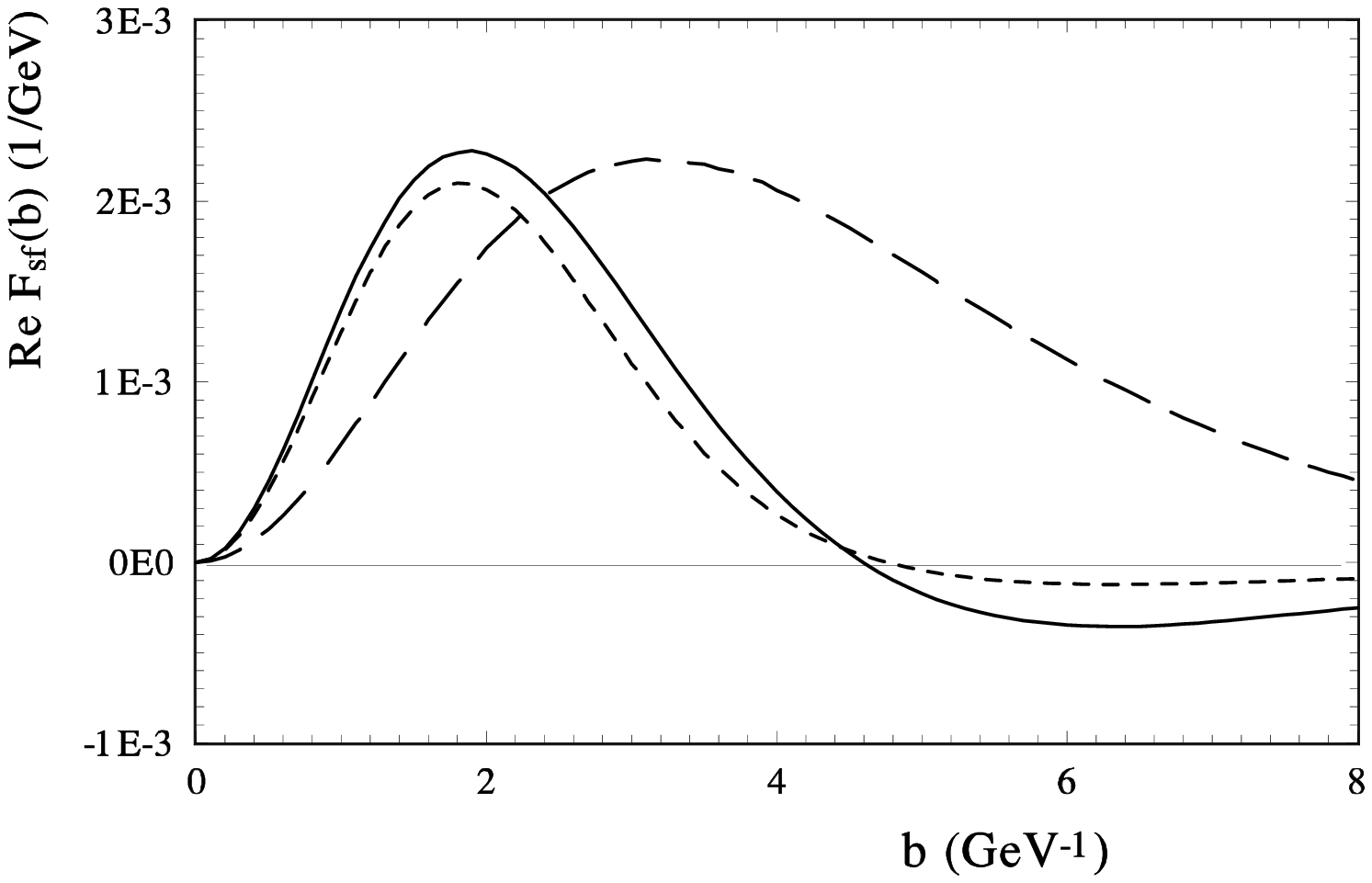} 
\caption{ The spin-flip amplitude in the impact parameter representation
(hard line - eq.(\ref{G-Kur}), dashed line -  eq.(\ref{Gauss})
  }
\label{Fig_3}
\end{figure}

\begin{figure}
\includegraphics[width=.45\textwidth]{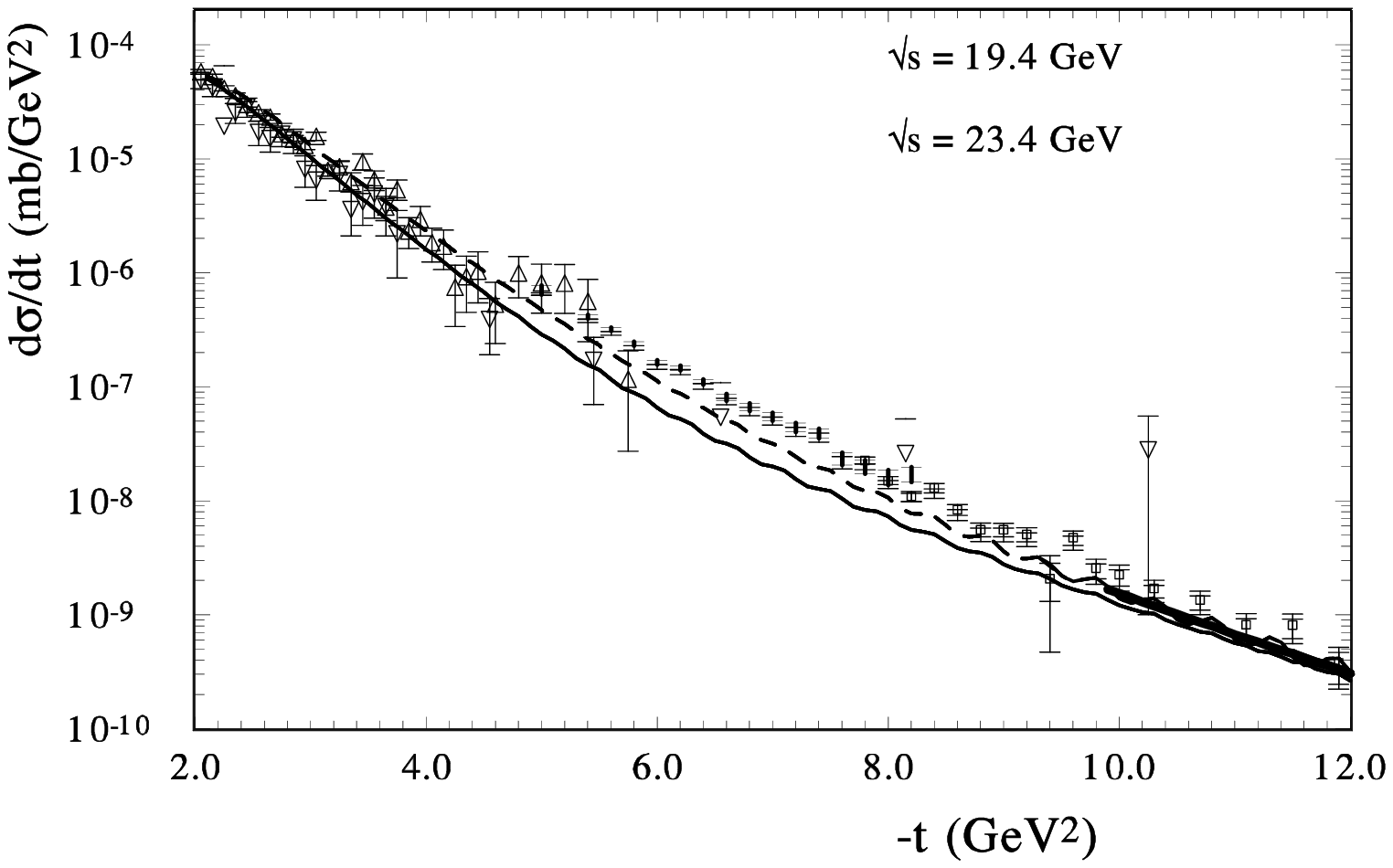} 
\includegraphics[width=.45\textwidth]{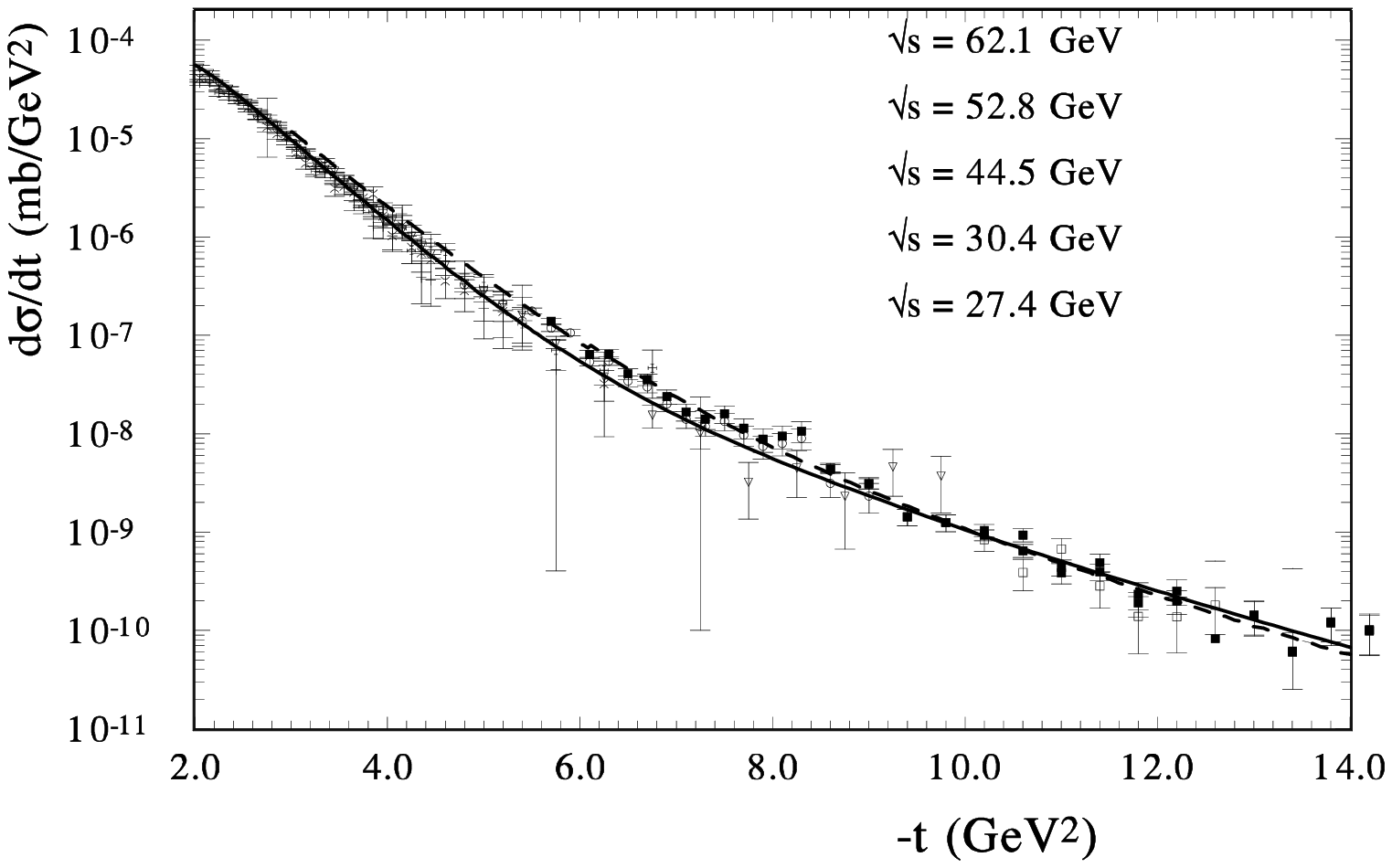} 
\caption{  Differential cross sections of the proton-proton elastic scattering
at large momentum transfer (hard line- the calculations at $\sqrt(s)=19.4$ GeV,
dashed line - the calculations at $\sqrt(s)=52.8$ GeV).
  }
\label{Fig_14}
\end{figure}

   In \cite{G-Kuraev1,G-Kuraev2} on the basis of  generalization of the constituent-counting rules of the perturbative QCD  the proton current matrix elements $J^{\pm\delta\delta}_{p}$
   for the full set of  spin combinations  corresponding to the number
   of the spin-flipped quarks was calculated. It  leads to part of the spin-flip amplitude
  \ba
              F_{h}^{sl}  \sim \sqrt{-t}/(\frac{4}{9} m_{p}^{2}) \  \sqrt{-t}/(\frac{4}{9} m_{p}^{2}) \
              \sqrt{-t}/(\frac{4}{9} m_{p}^{2}).
       \label{GKur-sflip}
\ea
  Hence, such an amplitude will  give  large contributions at large momentum transfer.
  In the HEGS model the calculations are extended up to $-t=15 $ GeV$^2$,
  and we added the small contribution of the asymptotically  independent  energy part of
   the spin-flip amplitude  with the kinematical factor  eq.(\ref{GKur-sflip}).
   So, the form of the spin-flip amplitude is determined as
  \begin{eqnarray}
  F_{sf}(s,t) \ =  h_{sf} q^3 F_{1}^{2}(t) e^{-B_{sf} q^{2}}. 
\label{G-Kur}
  \end{eqnarray}
   Of course, at  lower energy we need to take into account
  the energy dependence parts of the spin-flip amplitudes.
  However, it requires including in our examination of additional
  polarization data and the contributions of the non-leading Reggions, which essentially complicate the picture.  Now it is  beyond the scope of this paper.
  Such a contribution  can be made in future works.

    It is interesting that the spin-flip amplitude in the form (\ref{G-Kur})
    can be compared with the spin-flip amplitude with the standard kinematical factor $\sqrt{|t|}$
     but with  the  Gaussian form of the slope
   \begin{eqnarray}
  F_{sf}(s,t) \ =  h_{sf} q F_{1}^{2}(t) e^{-(r-q)^{2}}. 
\label{Gauss}
  \end{eqnarray}

\begin{figure}
  \includegraphics[width=.3\textwidth]{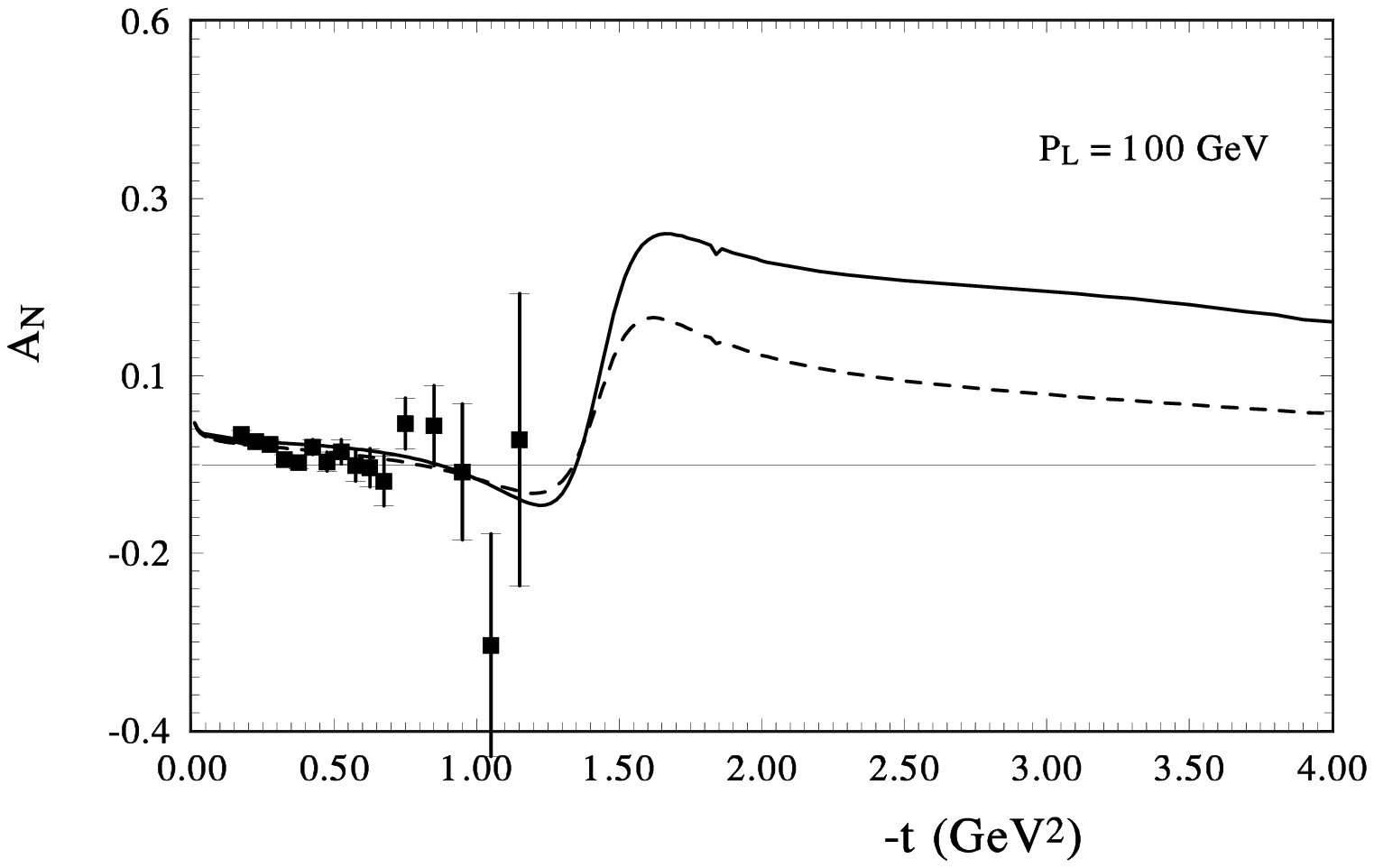}
      \includegraphics[width=.3\textwidth]{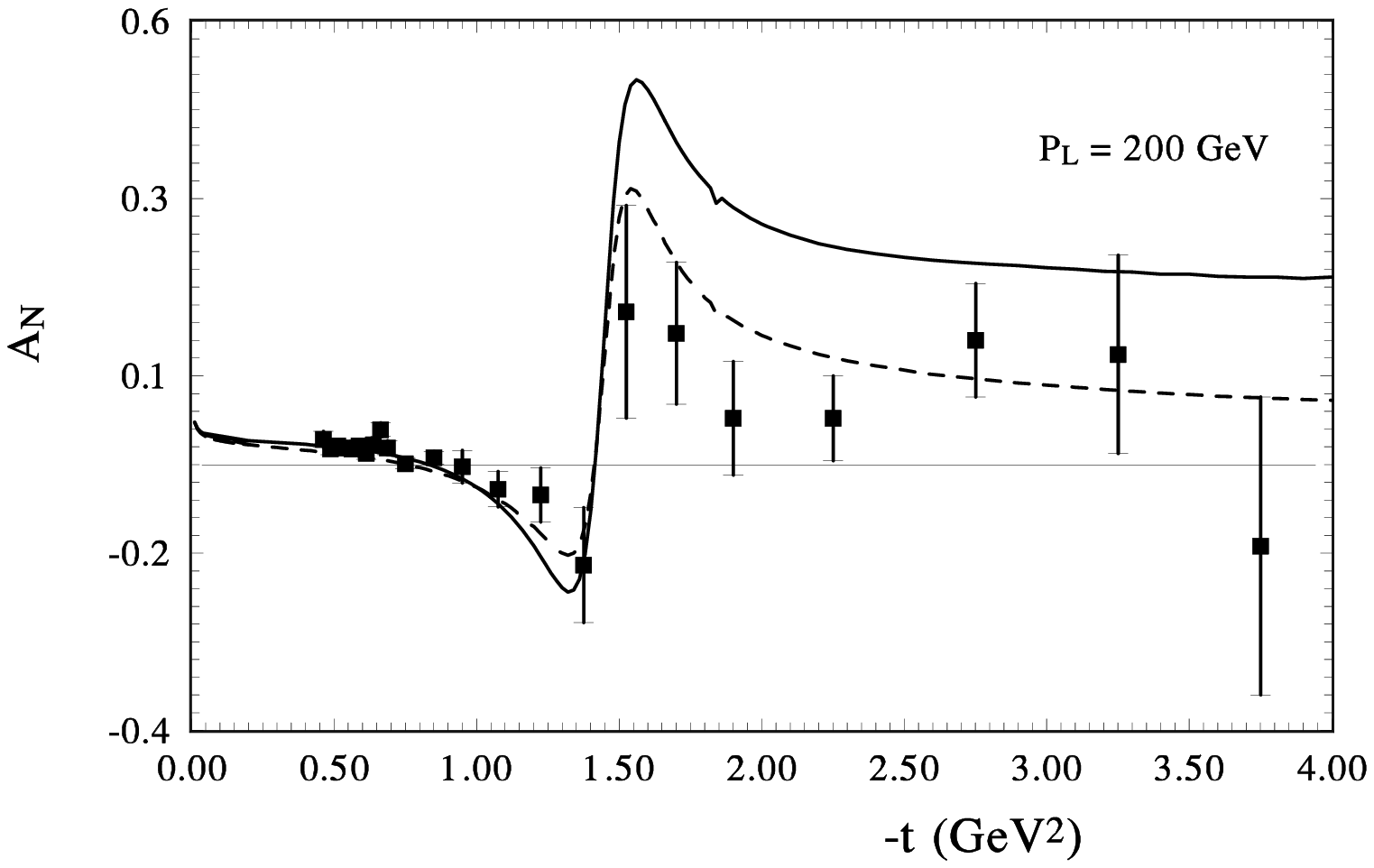}
            \includegraphics[width=.3\textwidth]{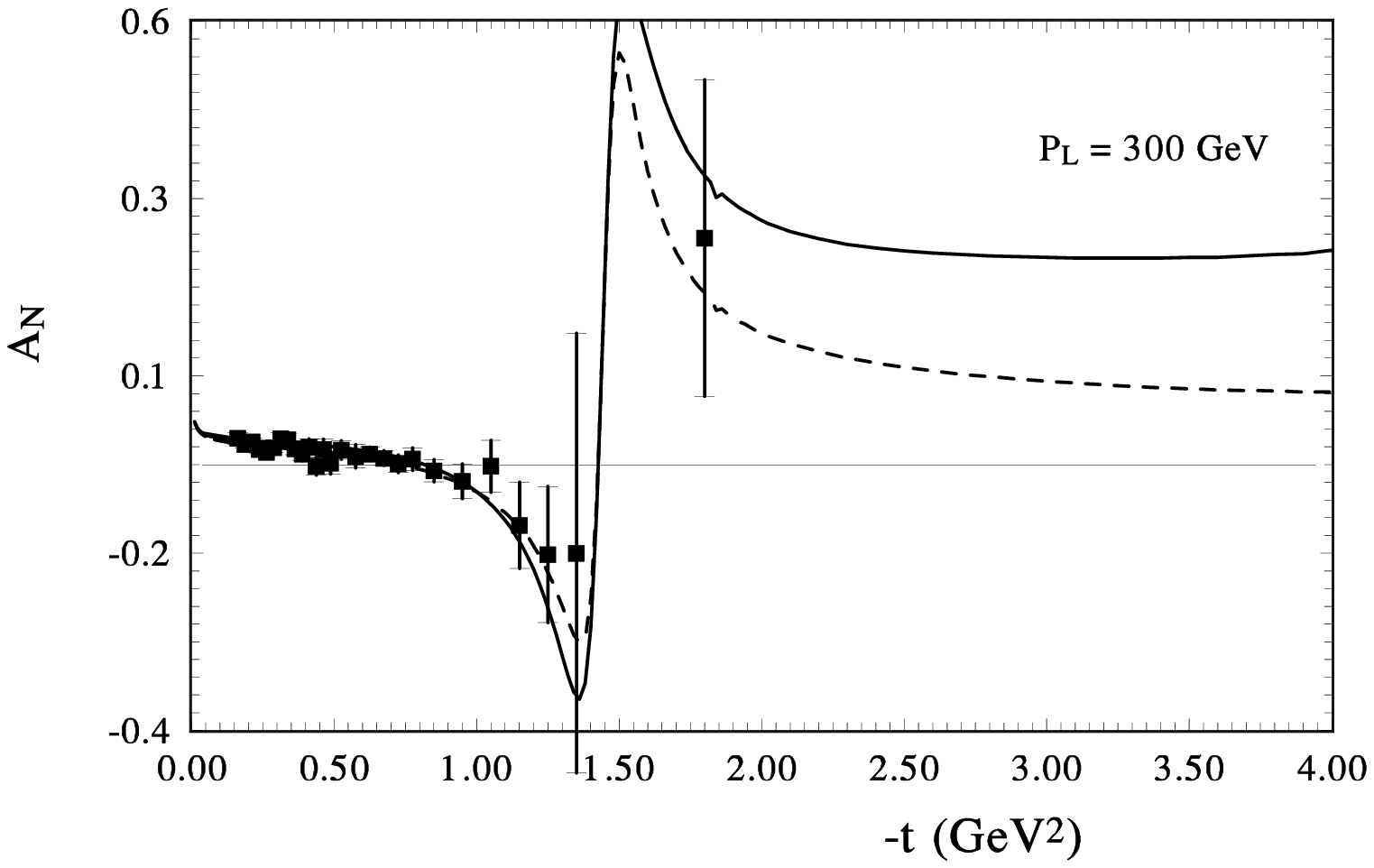}
  \caption{The spin correlation parameter $A_{N}(s,t)$ for the elastic proton proton scattering
  at $p_{L}=100$ GeV, $p_{L}=200$ GeV, and $p_{L}=300$ GeV.
  (hard line - model calculations with the spin-flip amplitude, eq.(\ref{G-Kur}),
  dashed line - the model calculations with the spin-flip amplitude, eq.(\ref{Gauss}),
  points - the experimental data \cite{200a,200b,300}.
   }
 \label{Fig5}
\end{figure}

        Let us compare the spin-flip amplitudes  (\ref{G-Kur}) and (\ref{Gauss}) in the impact parameter representation

 \begin{eqnarray}
  F_{sf}(b,t) \ = 
   \ \int_{0}^{\infty} \  q \ dq  J_{1}(q,b) F_{sf}(s,q) .
\label{tot0}
\end{eqnarray}

    The spin-flip amplitudes with the standard kinematical factor,  eq.(\ref{G-Kur}),
      and with the kinematical factor, eq.(\ref{GKur-sflip}), are
          compared with the spin-flip amplitude, eq.(\ref{Gauss}), on Fig.3a.
          It can be seen that the standard spin-flip amplitude has more peripheral interactions.

 A more obvious result can be obtained by comparing  of these amplitude multiplied
 by the impact parameter $b$  and with the same sizes ( we devide the standared amplitude by factor 2
 and multiply amplitude eq.(\ref{Gauss}) by the same factor).
 The results present in Fig.3b. The impact dependence of the  amplitudes, eq.(\ref{G-Kur}),
 and,  eq.(\ref{Gauss}),
  has a small difference.
 However, the impact dependence of the standard spin flip amplitude with the kinematical factor $\sqrt{|t|}$
 has an essentially different form. It has a more peripheral origin.

    As we have already noted that the size of the spin-flip contributions is bounded by the size
    of the differential cross sections at diffraction minimum of the differential cross sections.
    In fig.2, the model calculations are compared for the energy where the diffraction dip
    has the minimum value. A similar situation for the proton-proton cross sections occurs
    at $\sqrt{s}=30$ GeV where the real part change its sign at $t=0$. The model calculations
    describe the form of the diffraction minimum at this energy very well.
    Moreover, the model sufficiently well reproduces the energy dependence of the differential
    cross section in the region of the diffraction dip. Note that in the HEGS model
    the real part of the elastic scattering amplitude is determined by the complex $\hat{s}$ only.
    So  Fig.2 shows that the contribution of the spin-flip amplitude is essentially small
    in this region of the momentum transfer.
      However, the contributions of the taken form of the energy independent spin-flip amplitude, eq.(\ref{G-Kur}),
      allow one to describe the differential cross sections at large momentum transfer.
        The experimental data on the differential cross section of the elastic proton-proton scattering
       at large momentum transfer  show the small energy dependence (see Fig.4).
       The HEGS model calculations reproduce a such small energy dependence of the differential cross sections
       at large $t$ (see Fig.4).

   Now let us calculate  in the framework of the HEGS model the  spin correlation parameter $A_N(s,t)$
  in the region of the diffraction minimum.
     The comparison with  the existing experimental data  at higher energies,
     where the experimental data are available,
      $p_{L}=100$ GeV,  $p_{L}=200$ GeV, $p_{L}=300$ GeV are shown in Fig.5.
      In the last case, the diffraction minimum has a
  large dip. We can see that  both amplitudes, eq.(\ref{G-Kur}), and, eq.(\ref{Gauss}), give
  a similar picture and qualitatively agree with the existing experimental data.
      However, for a quantitative description more new more precise experimental data
      at high energies are needed.

  Our predictions for the spin correlation parameter $A_{N}$
  are presented in Fig.5 for two energies
  $\sqrt{s}=50$ GeV and  $\sqrt{s}=500$ GeV and for all three examined spin-flip amplitudes.
  We can see that the total behavior is practically the same for all spin-flip amplitudes.
 The  difference in the size of $A_{N}$  is presented  more remarkably  at large momentum transfer.

\begin{figure}
  \includegraphics[width=.45\textwidth]{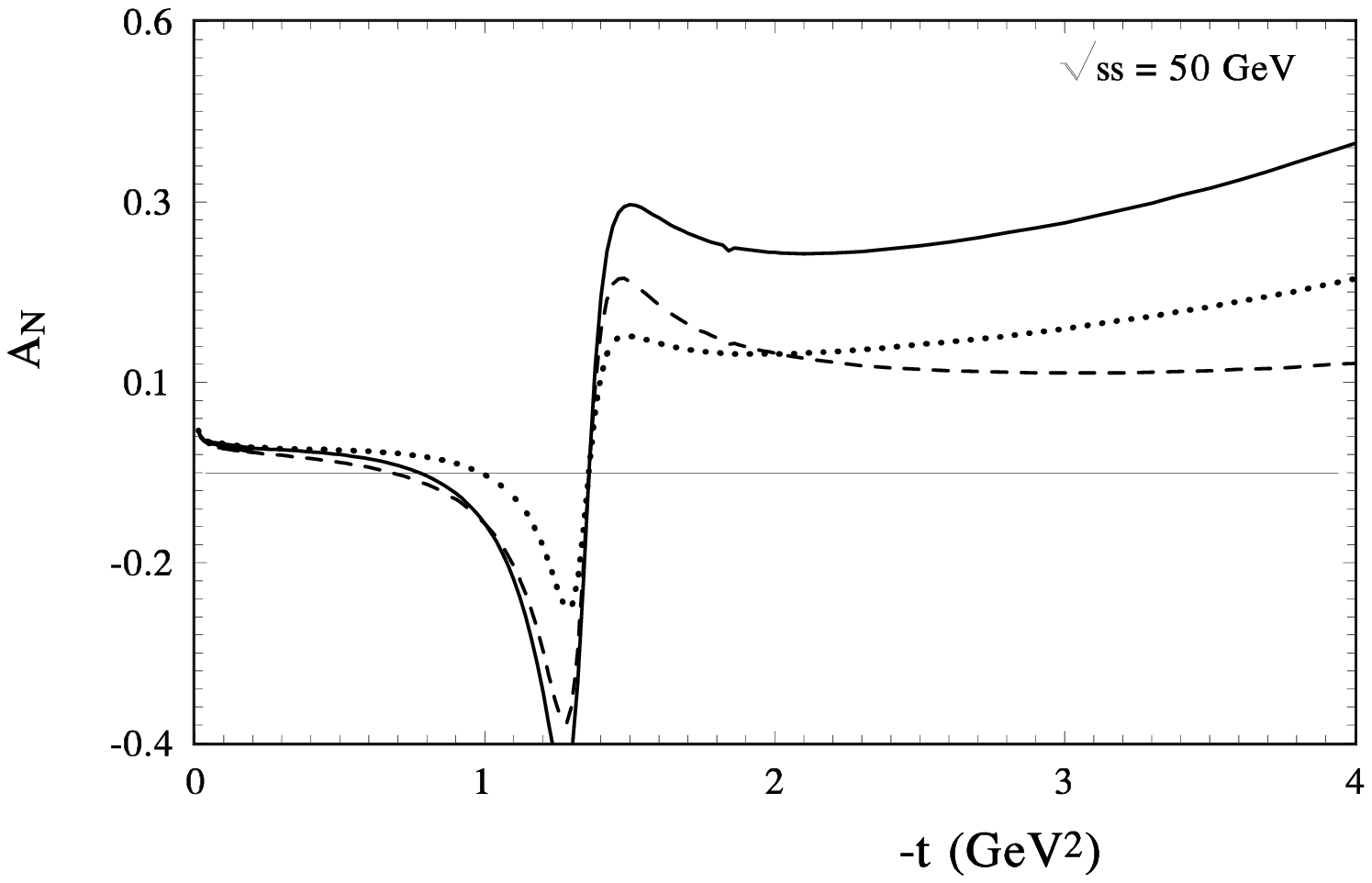}
      \includegraphics[width=.45\textwidth]{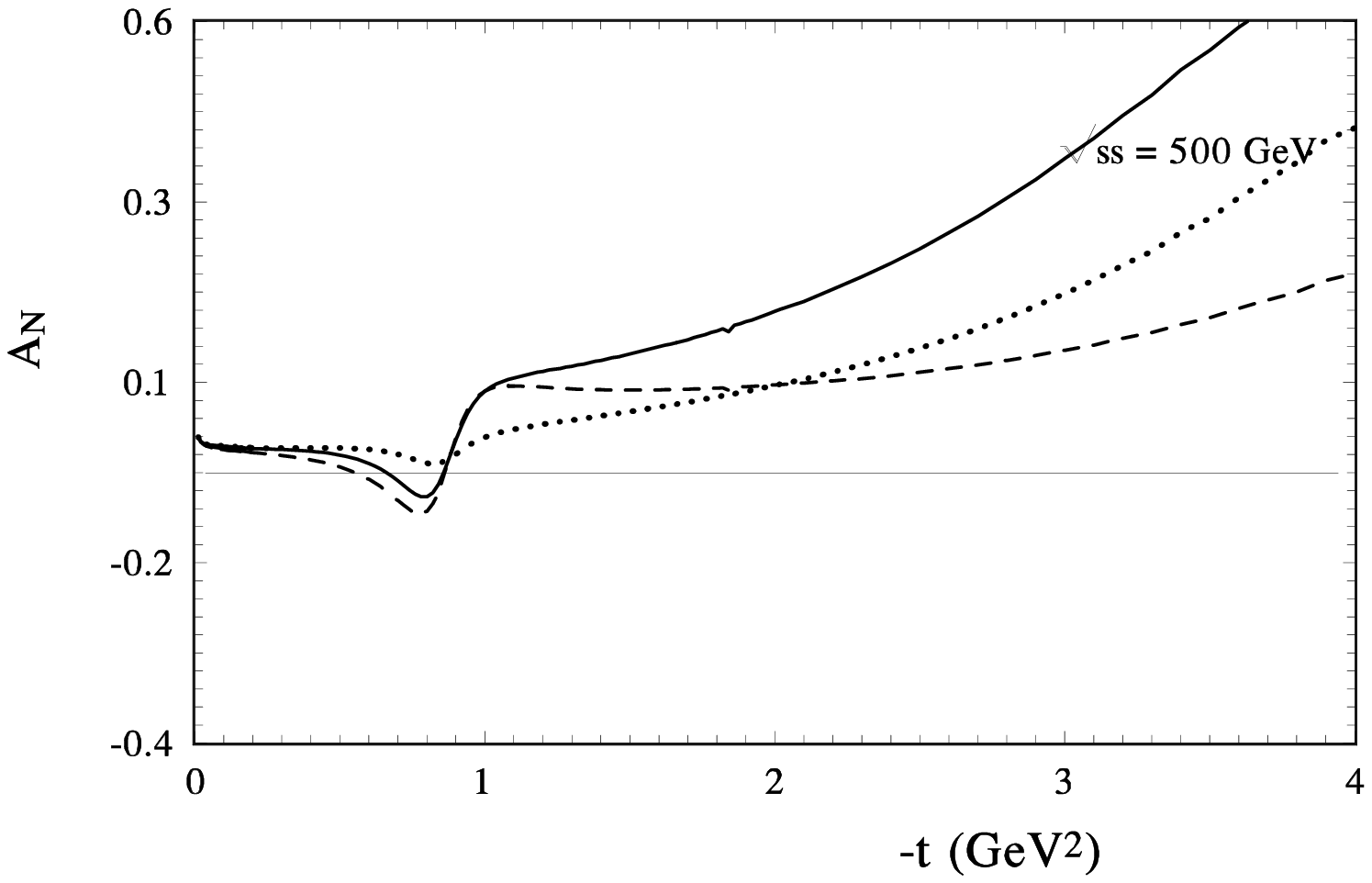}
  \caption{ The same as in Fig.5 (additional points line - the calculation with the spin-flip amplitude,
  eq.(\ref{G-Kur}, with the standard factor $\sqrt{|t|}$\\
  a) (right panel) at $\sqrt{s}=50$ GeV; b) (left panel) at $\sqrt{s}=500$ GeV
   }
 \label{Fig5}
\end{figure}


    \section{Conclusion}

  The existing experimental data on the elastic hadron scattering at high energies and large momentum transfer show a small energy dependence of the differential cross sections.
  In the framework of the Donnachie-Landshoff model \cite{DL} such an effect is explained by the energy independent Odderon contribution.
  In our HEGS model , which describes the maximum experimental data in a wide energy region
  from $\sqrt{s}=9$ GeV  $\sqrt{s}=8$ TeV
  with minimum fitting parameters, the odderon has the same intercept as the standard soft Pomeron.
    Taking into account the form of the spin-flip amplitude, proposed
    in the works of Galynskii-Kuraev \cite{G-Kuraev1,G-Kuraev2},
    the energy independence  of the differential cross sections at large momentum transfer is explained
     by the contribution of the energy independent part of the spin-flip amplitude.
    The obtained size and momentum transfer dependence of the spin-flip amplitude allow one to describe the differential cross sections in the region of the diffraction minimum and at large momentum transfer.

      We show that such a spin-flip amplitude is related to the interaction at a small radius in the hadron.
        It is found, that such behavior can be modeled by the spin-flip amplitude with the standard
        kinematical factor proportional to $\sqrt{|t|}$, but with the Gaussian form of the slope.
        The spin correlation parameter $A_{N}(s,t)$  calculated in the framework of the model does not
        contradict the existing experimental data at high energies. We hope that the forward experiments
        at a future accelerator can give  valuable information for the improvement of our theoretical
        understanding of the spin-dependent hadron interactions. This is especially true for  future experiments at NICA with a polarized beam and target.

\vspace{0.5cm}
{\bf Acknowledgments}
 {\it The authors would like to thank J.-R. Cudell and O.V. Teryaev
   for fruitful   discussions of some questions   considered in the paper.} \\

  \end{document}